\pdfoutput=1
\documentclass[fleqn,10pt]{wlscirep}
\usepackage[utf8]{inputenc}
\usepackage[T1]{fontenc}
\usepackage{amssymb,amsmath,nicefrac}

\usepackage[color,draft]{showkeys}

\title{Mechanical response of dense pedestrian crowds to the crossing of intruders}

\author[1,*]{Alexandre NICOLAS}
\author[2]{Marcelo KUPERMAN}
\author[2,3]{Santiago IBA\~NEZ}
\author[2]{Sebasti\'an BOUZAT}
\author[4]{C\'ecile APPERT-ROLLAND}
\affil[1]{LPTMS UMR 8626, CNRS, Universit\'e Paris-Sud, Universit\'e Paris-Saclay, 91405 Orsay, France.}
\affil[2]{Consejo Nacional de Investigaciones Científicas y Técnicas, Centro Atómico Bariloche (CNEA) and Instituto Balseiro,
R8400AGP Bariloche, Argentina.}
\affil[3]{Universidad Nacional de R\'io Negro, Sede Andina, 8400 Bariloche, Argentina.}
\affil[4]{LPT, CNRS UMR 8627, Universit\'e Paris-Sud, Universit\'e Paris-Saclay, 91405 Orsay, France.}

\affil[*]{alexandre.nicolas@polytechnique.edu}

\keywords{Mechanics of disordered media, pedestrian crowds, granular media.}

\begin{abstract}
The increasing number of mass events involving large crowds calls for a better
understanding of the dynamics of dense crowds. Inquiring into the possibility of 
a mechanical description of these dynamics, we experimentally study the crossing of dense static crowds
by a cylindrical intruder, a mechanical test which is classical for granular matter.

The analysis of our experiments reveals robust features in the crowds' response, comprising both similarities and discrepancies with the response of granular media.
Common features include the presence of a depleted region behind the intruder and the short-range character of the perturbation.
On the other hand, unlike grains, pedestrians anticipate the intruder's passage by moving much before contact
and their displacements are mostly lateral, hence
not aligned with the forces exerted by the intruder. Similar conclusions
are reached when the intruder is not a cylinder, but a single crossing pedestrian.

Thus, our work shows that pedestrian interactions even at high densities (3 to 6 ped/m$^2$) do not reduce to mechanical ones.
More generally, the avoidance strategies evidenced by our findings question the incautious use of 
force models for dense crowds.

\end{abstract}
\begin{document}

\flushbottom
\maketitle

\thispagestyle{empty}

\section*{Introduction}

Magritte's surrealistic painting \emph{Golconda} depicts a suburban landscape with scores of passers-by stiffly standing in mid-air, as if they were `raining' over the city. On close inspection the individuals are dissimilar, but overall they look nearly identical, as though being part of a large group rendered their differences as faint as those between rain droplets. This analogy between pedestrians and beads would considerably facilitate the prediction of pedestrian crowd flows, if one were to take it at face value.

Predictions for pedestrian flows are of great avail for the architectural design of large facilities and for the management of throngs, whether it be in underground stations\cite{hoogendoorn2004applying}, at mass religious events\cite{helbing2007dynamics} or sports gatherings\cite{bain_b2018}, or in front of a shop just before its opening on Black Fridays in the United States. The fear of stampedes in such assemblies has turned much of the research effort on dense crowds towards extreme situations, such as past crowd disasters \cite{helbing2007dynamics, krausz2012loveparade} or scenarios of evacuation through a door \cite{hoogendoorn2005pedestrian,faure2015crowd,garcimartin2016flow, nicolas2017pedestrian}, but at high densities complex collective phenomena are also expected in ordinary situations and require proper modelling. Nowadays, 
pedestrian simulation software is routinely entrusted with the task of predicting
the dynamics of crowds. Nevertheless, their underlying models still lack extensive validation, especially at very high densities. Admittedly,
the outputs of simulations made with commercial software have been compared to
some high-density real-life situations
\cite{berrou2007calibration,dridi2014pedestrian,isenhour2014verification},
but such comparisons are generally restricted to macroscopic observables - for example evacuation times - and visual impressions. In this vein, it is symptomatic that Fruin's Levels of Service\cite{fruin1971pedestrian}, widely used in safety
and design handbooks, amalgamate all densities above 2.15~ped/m$^2$ into Level F, without further distinction. Accordingly, dense pedestrian crowds in generic settings deserve more attention than they have received so far \cite{haghani2018crowd}.

As a matter of fact, beyond Magritte's painting, analogies between the dynamics of these crowds and the mechanics of natural systems, in particular granular media, have long been hinted at \cite{helbing1998similarities}, and have now been partly confirmed\cite{helbing2007dynamics,ma2013new,Zuriguel2014clogging,Pastor2015experimental}. The comparisons concern the passage of a bottleneck~\cite{Zuriguel2014clogging,Pastor2015experimental} and the onset of
turbulence~\cite{helbing2007dynamics,ma2013new}. Inspiration could thus be drawn from these scientifically riper fields, where
progress was made by identifying the elementary microscopic processes by which the material deforms (motion of dislocations in crystals, shear transformations in amorphous
solids)
as well as the response halo generated by these elementary perturbations. Along these lines,
since crowds are often traversed by intruders (pedestrians, bicycles, vehicles, etc.), it would be valuable
to know the crowd's response to such crossings. It so happens that intruding a cylinder 
into a medium is actually a classical mechanical test. In particular, the response to such a perturbation
strongly discriminates between granular media and viscous fluids, with
an exponential radial decay of the perturbation in the former case \cite{seguin2013experimental,kolb2014flow} and a very long-ranged
impact in the latter case \cite{kaplun1957low}. 

In this paper, we inspect the response of a dense static crowd induced by the crossing of an intruder. One may think that steric constraints will bring the crowd's response close to that of a granular medium, but an exciting question is how the pedestrians' decision-making processes will couple with the strong mechanical constraints.
More generally, we aim to (i) collect controlled experimental data on dense crowds  that can be used to test and calibrate models, (ii) explore the similarities of the crowd's response to the intrusion of a cylinder with that of granular media, and (iii) discuss the case in which the intruder is a single pedestrian, a situation often encountered in daily-life. In so doing, we lay the first stone for the empirical development of the (continuum) mechanics of dense crowds.

\section*{Presentation of the experiments}

In order to study the response of a static crowd to the passage of an intruder, two
series of experiments were performed in gymnasiums in Orsay (France) and Bariloche (Argentina) in June 2017 and September 2017, respectively. They involved between
35 and 40 voluntary participants (men and women), aged 20-60, mostly students.

\begin{figure}[ht]
\centering
\includegraphics[width=\linewidth]{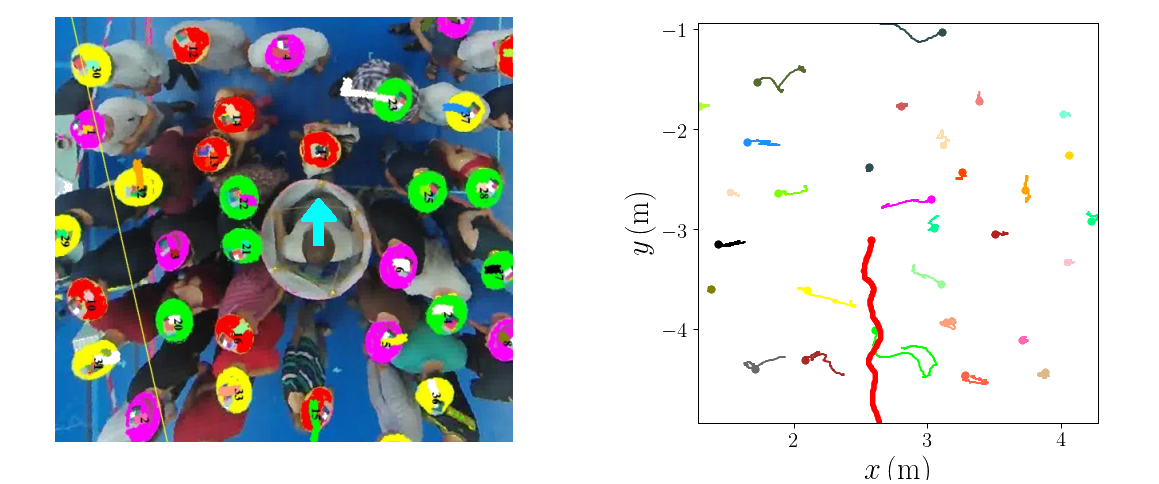}
\caption{Crossing of a fairly dense crowd ($\bar\rho \approx 3.5 \mathrm{ped/m^2}$)
by a cylindrical intruder. \textbf{(a)} Snapshot of an experiment (France), after
automatic detection and tracking of the participants' hats. \textbf{(b)} Trajectories of the intruder (thick red line) and of the participants over a time window of around 5 seconds.
}
\label{fig:figure1}
\end{figure}

During the experiments, the
participants stood in a delimited area, which allowed us to control the global density $\bar\rho$ (between 2 and 6 $\mathrm{ped}/\mathrm{m}^2$)
This static crowd was successively crossed from end to end [along the $y$-axis, see
Fig.~\ref{fig:figure1}(b)] by (i) an intruder of cylindrical shape
(diameter: 74~cm in France, 68~cm in Argentina) moved along a more or less straight line $x=\mathrm{cst}$ by
a staff member located inside it [see Fig.~\ref{fig:figure1}(a)], or (ii) a randomly
chosen participant, asked to make his/her way to the other side. Meanwhile,
people in the standing crowd either faced the incoming intruder (France and
Argentina), or were prescribed random orientations (France), or turned their
backs to the crowd (Argentina). Furthermore, we gave distinct instructions
in France and in Argentina: In France, the participants were asked to
behave casually, as if they were standing on an underground platform , whereas
in Argentina they were to make way for the intruder only upon coming in contact
with it.

The experiments were recorded from above, at an altitude of about 4 meters, with high-resolution 60~Hz action cameras (Tomtom Bandit in France, GoPro Hero 3 in Argentina). After correction of the optical distortion of the cameras, home-made software allowed us to detect and track the coloured hats worn by the participants, as shown in Fig.~\ref{fig:figure1}. The extracted data were then post-processed manually. As detailed in Ref.~\cite{appert2018experimental}, we found an absolute experimental uncertainty
of the order of 10 to 15~cm in most cases.
The uncertainty cannot be reduced much below,
unless one corrects for the different participants' heights or one positions the camera extremely high \cite{Boltes2016influences}.
Note, however, that the error concerning the displacement of a \emph{given}
participant will be much lower than that on absolute positions. More extensive manual post-processing
was required for the experiments conducted in Argentina, because red headscarves with a white mark were used there instead of hats of various colours.

\section*{Results}

\subsection*{Crowd's response to a standardized perturbation}

This section addresses the following questions:
What are the main features of the crowd's response to its crossing by a cylindrical intruder? How similar is it to what has been reported for two-dimensional granular media \cite{kolb2014flow,seguin2011dense}?

\subsubsection*{Density field around the intruder}
\begin{figure}[ht]
\centering
\includegraphics[width=\linewidth]{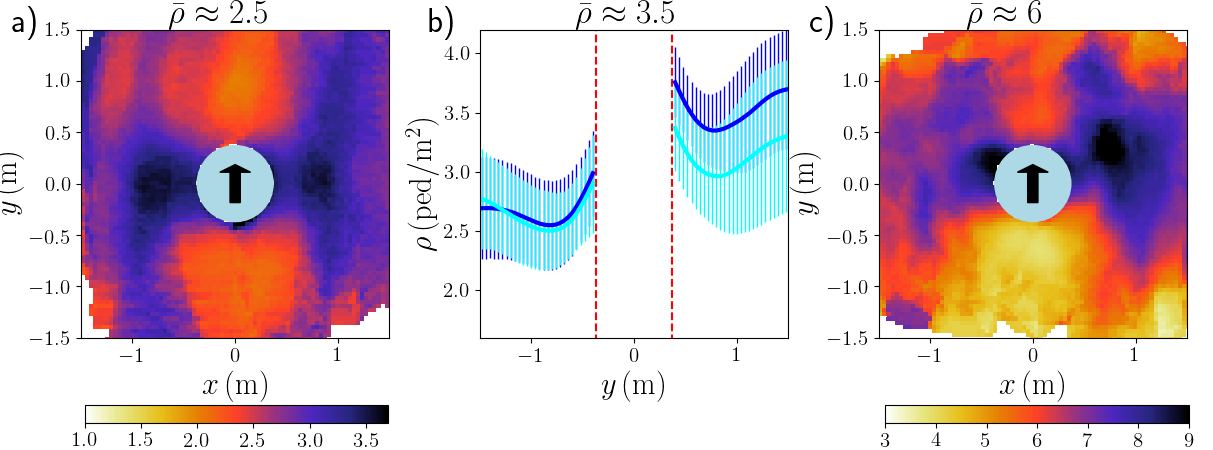}
\caption{Pedestrian densities (in $\mathrm{ped}/\mathrm{m}^2$) in the frame of the moving cylindrical intruder, in crowds \emph{facing the intruder} (France). 
Panels \textbf{(a)} and \textbf{(c)} show the average density fields  in moderately dense crowds ($\bar\rho \approx 2.5$, average over 7 crossings, which we abbreviate into `x7')
and very dense crowds ($\bar\rho \approx 6$, x7), respectively.
Panel \textbf{(b)} displays the density profiles along the `streamline' $x=0$ in fairly dense crowds ($\bar\rho \approx 3.5$); the result for
participants facing the intruder (light blue, x9) is compared with its counterpart for randomly oriented participants (dark blue, x9). The error bars represent the standard deviation over all similar  crossings and the dashed red lines delimit the portion of space occupied by the intruder.
}
\label{fig:figure2}
\end{figure}

First, we consider the density fields around the intruder, averaged over time and over
similar realisations in the intruder's frame and plotted in Fig.~\ref{fig:figure2} for crowds facing the intruder. The local densities were computed with a novel method (detailed in the \emph{Methods} section), which is based on
a Voronoi tessellation of the plane, but suitably modified to account for edge effects.

\begin{figure}[ht]
\centering
\includegraphics[width=0.75\linewidth]{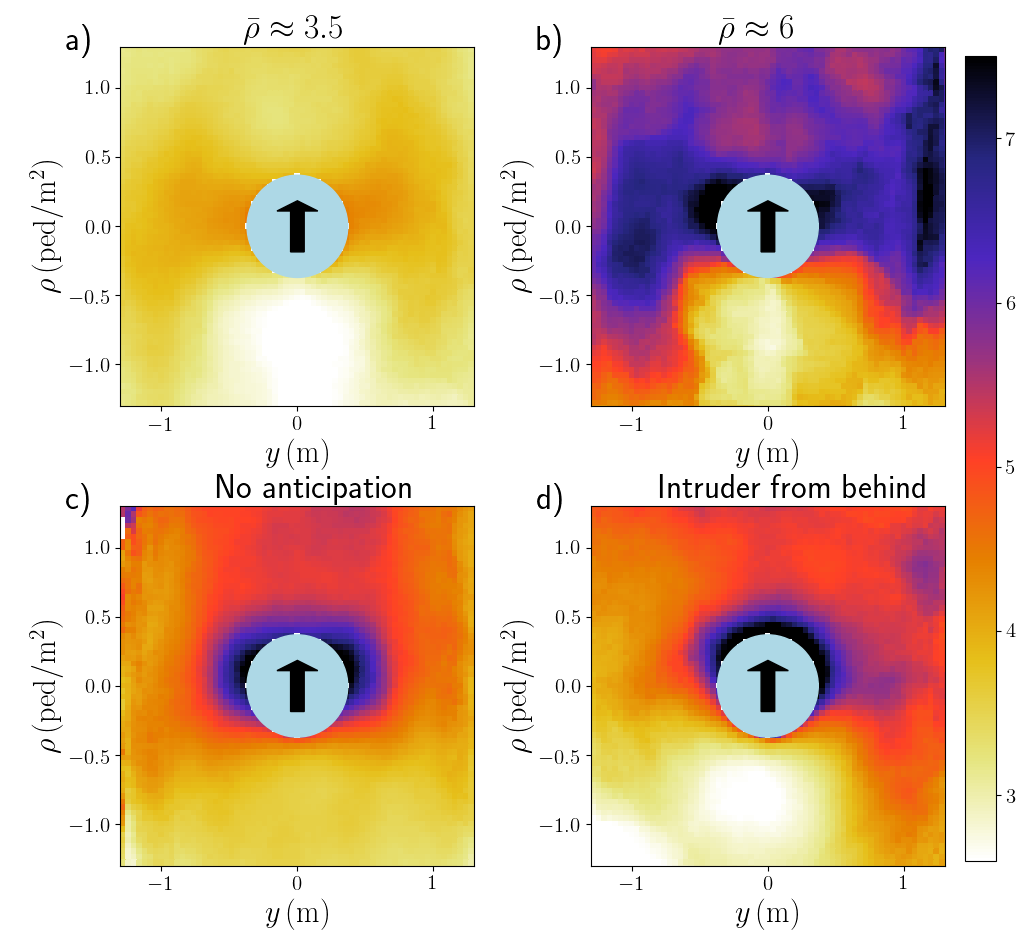}
\caption{Average density fields (in $\mathrm{ped/m^2}$) around the cylindrical intruder for various types of dense crowds.
\textbf{(a)} Fairly dense crowd ($\bar\rho \approx 3.5$) with randomly oriented participants (France, x9).
\textbf{(b)} Very dense crowd ($\bar\rho \approx 6$) with randomly oriented participants (France, x5).
\textbf{(c)} Dense crowd ($\bar\rho \approx 4.5$) with non-anticipating participants facing the intruder (Argentina, x15).
\textbf{(d)} Dense crowd ($\bar\rho \approx 4.5$) with participants turning their backs to the intruder (Argentina, x10).}
\label{fig:figure3}
\end{figure}

In all situations, a depleted region behind the cylinder clearly stands out in the density maps. It reflects the empty space created in the intruder's wake, which is then filled by pedestrians in a matter of seconds, similarly to the cavity formed downstream from the intruder in granular media \cite{kolb2013rigid,kolb2014flow}. However, it is remarkable that, in contrast with granular matter (below jamming), compaction is observed mostly along the transverse direction: high-density `wings' adorn both sides of the cylinder. We emphasise that these features are witnessed at all densities, with participants facing the intruder [Fig.~\ref{fig:figure2}] as well as randomly oriented ones [Fig.~\ref{fig:figure3}].

Upstream from the cylinder, 
instead of the compression reported in granular matter below jamming~\cite{kolb2014flow}, 
we observe a redistribution of the density, with higher values on the sides, and either lower or average density along the central axis
[Figs.~\ref{fig:figure2}-\ref{fig:figure3}].
This can be ascribed to the pedestrians' dodging out of the intruder's trajectory with anticipation, as one clearly sees in the videos. 
Consistently with this idea, the effect is more marked when pedestrians face the intruder than for random orientation, which also transpires from the density profiles
along the longitudinal axis $x=0$, presented in Fig.~\ref{fig:figure2}(b) for moderately dense crowds
(the local densities plotted in the density profiles were smoothed over a length scale $\xi=9\,\mathrm{cm}$ according to the procedure described in the \emph{Methods}).  

When pedestrians are told not to anticipate (in Argentina),
the density field becomes more reminiscent of (but probably not as marked as)
the granular case, with some compaction upstream.
It should be noted that in spite of the instructions, some pedestrians
facing the intruder [Fig.~\ref{fig:figure3}(c)] still had a tendency
to anticipate, visible in the videos.
Only when they were turning their back to the intruder
was anticipation fully suppressed.
As a result, it is only in this case that  a clear
high-density layer coats the fore part of the intruder [see Fig.~\ref{fig:figure3}(d)].

\subsubsection*{Velocity fields around the intruder}

To shed light on the emergence of the density inhomogeneities, we take a closer
look at the pedestrians' displacements. To this end, we measure the velocity fields $\boldsymbol{v}(x,y)= \delta \boldsymbol{u}(x,y) / \delta
t$ around the intruder.
Here $ \delta \boldsymbol{u}(x,y)$ is the displacement between $t$ and $t+\delta t$ of a pedestrian standing at position $(x,y)$ (relative to the intruder) at time $t$.
Continuous fields are obtained via the smoothing procedure described
in \emph{Methods}.

\begin{figure}[ht]
\centering
\includegraphics[width=\linewidth]{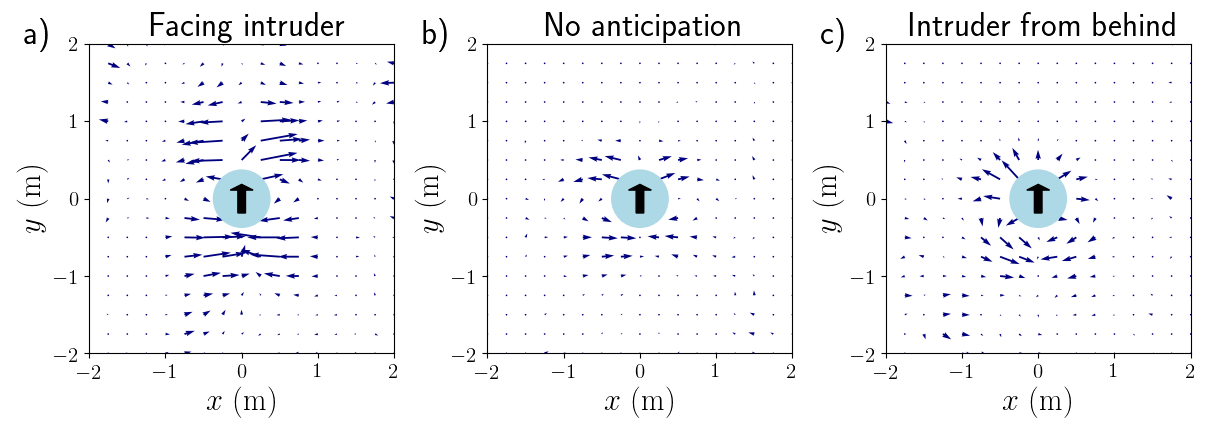}
\caption{Mean velocity fields around the cylindrical intruder.
\textbf{(a)} For a fairly dense crowd ($\bar\rho \approx 3.5\,\mathrm{ped/m^2}$) with participants facing the intruder (France, x9).
\textbf{(b)} For a dense crowd ($\bar\rho \approx 4.5\,\mathrm{ped/m^2}$) with non-anticipating participants facing the intruder (Argentina, x15).
\textbf{(c)} For a dense crowd ($\bar\rho \approx 4.5\,\mathrm{ped/m^2}$) with non-anticipating participants turning their backs to the intruder (Argentina, x9). 
The velocities
are computed on the basis of the displacements performed over a time interval $\delta t=0.7\,\mathrm{s}$ and the arrows are magnified by a factor 3.
}
\label{fig:figure4}
\end{figure}

Let us first consider a crowd facing the intruder (France). 
Figure~\ref{fig:figure4}(a) presents the mean velocity field measured for  $\bar\rho \approx 3.5$ ped/m$^2$.
Interestingly, the velocities are almost exclusively
directed transversally (along $x$), i.e., perpendicularly to the trajectory of
the intruder: Ahead of it, the pedestrians step aside (towards larger
$|x|$) to avoid the anticipated contacts, while behind it they reoccupy the depleted region
(small $|x|$), again via lateral moves.

Very similar flow patterns are
observed when the density is varied, although the displacement amplitude
decreases with the density, and seems to extend farther transversally in the densest crowd [Supplemental Figure S1(a,b,c)]. 
The observed
overall similarity is noteworthy, because it spans across the transition from a
regime dominated by `social' interactions at low density to a tightly packed
one with prominent physical contacts.

When participants are randomly oriented [Supplemental Figure S1(d,e,f)],
the mean directions of motion remain
identical, i.e. transverse outwards (transverse inwards) in front of (behind)
the cylinder. However, the perturbed region ahead of it shrinks significantly, 
reflecting a lesser degree of anticipation, as the participants do not all face
the intruder. This shrinkage is even more marked when the participants were
asked not to anticipate the contact [Fig.~\ref{fig:figure4}(b)], in which case
the velocity field (whose overall pattern remains similar) localises closer
to the intruder.

The contrast sharpens with participants turning their backs to the intruder [Fig.~\ref{fig:figure4}(c)]. In this case, the velocity field, still localised in the direct vicinity of the cylinder, is no longer transverse, but mostly radial (at least, in the half-plane $y>0$). 
Besides, behind the cylinder, the velocities seem to acquire a small backwards longitudinal component, as though the participants, now that they can see the cylinder, wanted to stand away from it.
In fact, this flow pattern is strongly reminiscent of that observed around an intruder in a granular medium, shown in Supplementary Fig.~S2; the displacements are then probably
due to mechanical pushes by the intruder, without any preliminary avoidance manoeuvres. The wake of the intruder ($y<0$) is still reoccupied via lateral moves, but the latter are less unidirectional than in the other situations.

\subsubsection*{Spatial extent of the perturbation along the transverse direction}

We have noticed that the regions of highest density extend on both lateral sides of the intruder, while the velocity perturbation
seems to decay very quickly along the transverse direction $x$.
To quantify how much each pedestrian is affected by the intruder's passage on the whole, 
we now measure the amplitude $\boldsymbol {\Delta u^{(j)}}$ of the total displacement
of pedestrian $j$ defined by
\begin{equation}
\Delta u_x^{(j)} = \max_{t\in [t_i,t_f]} x^{(j)}(t) - \min_{t\in [t_i,t_f]} x^{(j)}(t)\text{ and }\Delta u_y^{(j)} = \max_{t\in [t_i,t_f]} y^{(j)}(t) - \min_{t\in [t_i,t_f]} y^{(j)}(t),
\label{eq:deltau}
\end{equation}
where times $t_i$ and $t_f$ mark the start and the end of the intruder's passage, respectively.

\begin{figure}[h]
\centering
\includegraphics[width=0.9\linewidth]{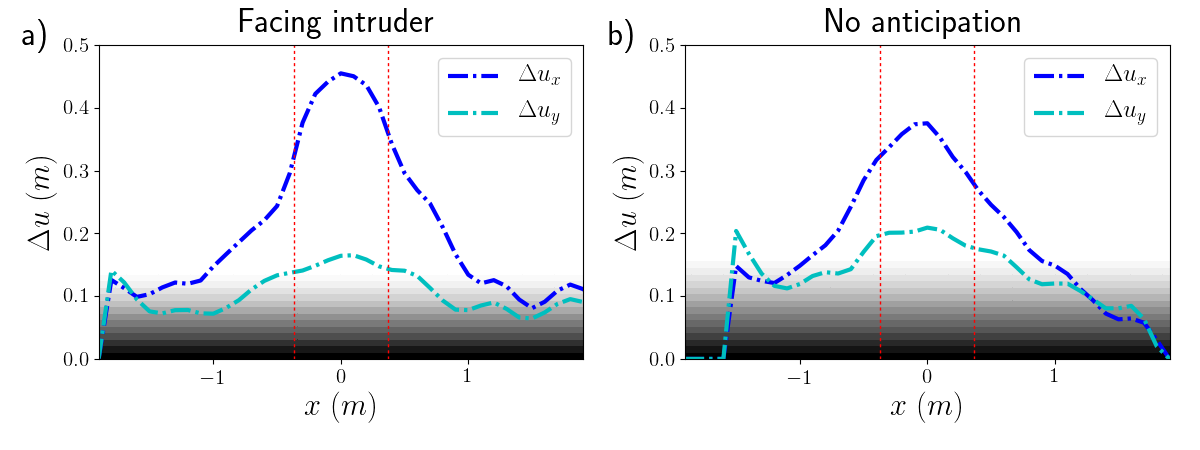}
\caption{Smoothed overall displacement amplitudes $\boldsymbol{\Delta u}$ induced by the cylindrical intruder, for 
\textbf{(a)} fairly dense crowds ($\bar{\rho} \approx 3.5\,\mathrm{ped/m^2}$) of participants facing the intruder and
\textbf{(b)} dense crowds ($\bar{\rho} \approx 4.5\,\mathrm{ped/m^2}$) of non-anticipating participants facing the intruder.
The grey-to-white overlay gives a rough idea of the background noise, due to the pedestrians' residual motion in the absence of a perturbation
and (to a lesser extent) to measurement errors.
The dashed red lines delimit the portion occupied by the intruder.
}
\label{fig:figure5}
\end{figure}

In Fig.~\ref{fig:figure5}, after  smoothing (see \emph{Methods}),
we plot  the variation of $\boldsymbol {\Delta u}$ with respect to the transverse distance between the initial pedestrian's
position $\left(x(t_i),y(t_i)\right)$  and the intruder's trajectory $\left(x_{cyl}(t),y_{cyl}(t)\right)$, i.e.,
$x(t_i)-x_{cyl}(t_c)$ where $t_c$ is such that 
$y_{cyl}(t_c)=y(t_i)$.

Consistently with our foregoing observations, we notice that the longitudinal displacement amplitude $\Delta u_y$ is much smaller than the transverse one $\Delta u_x$, especially in not too dense crowds.  As expected, the peak amplitude $\Delta u_x^\star$ slightly exceeds the cylinder's radius. Besides, 
the amplitude $\Delta u_x$ exhibits a fast decay in the transverse direction $x$ and 
drops to half of the peak value over a typical length scale $x_c\simeq D$, 
i.e., typically one radius beyond the outer boundary of the cylinder (note that experimentally the crowd was larger than $5\,x_c$ in width). This length scale does not vary much with the density and the experimental conditions, insofar as $x_c$ always ranges between $0.8\,D$ and $1.5\,D$ (perhaps excepting the experiments in which participants turned their backs to the intruder, for which $x_c$ could not be determined precisely).

\subsection*{Perturbation induced by the crossing of a single pedestrian}

\begin{figure}[ht]
\centering
\includegraphics[width=0.75\linewidth]{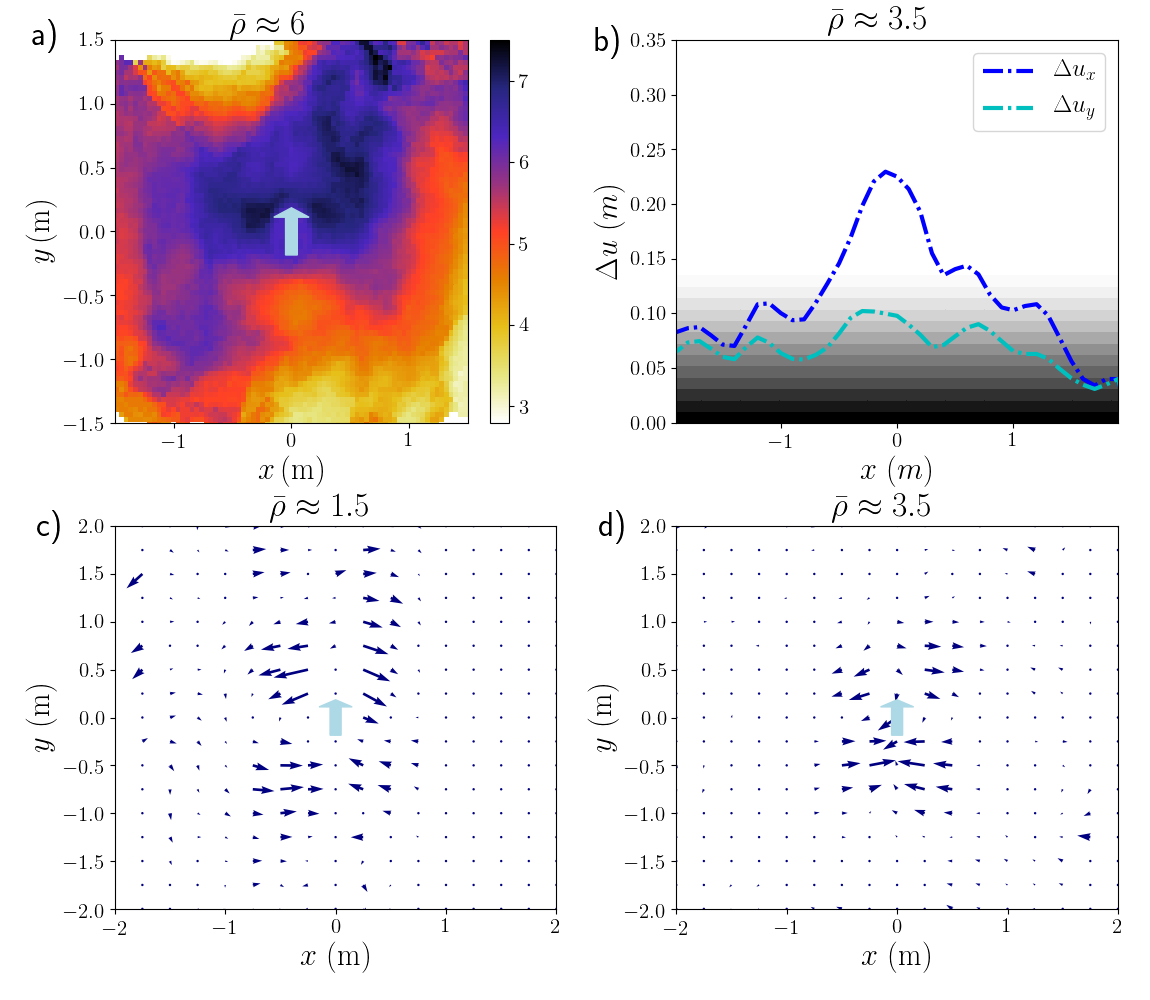}
\caption{Crowd's response to its crossing by a pedestrian.
\textbf{(a)} Average density field around the crossing pedestrian in a very dense crowd  ($\bar{\rho} \approx 6\,\mathrm{ped/m^2}$) of participants facing the intruder
(France, x10).
\textbf{(b)} Smoothed overall displacement amplitude $\Delta u$ in a fairly dense crowd  ($\bar{\rho} \approx 3.5\,\mathrm{ped/m^2}$) of participants facing the intruder (France, x10).
\textbf{(c-d)} Mean flow fields (based on the displacements over a time window $\delta t=0.7\,\mathrm{s}$) around the crossing pedestrian in crowds of participants facing the intruder at \textbf{(c)} fairly low density ($\bar{\rho} \approx 1.5\,\mathrm{ped/m^2}$) and \textbf{(d)} fairly high density ($\bar{\rho} \approx 3.5\,\mathrm{ped/m^2}$) (France, x5 and x8, respectively). The velocity arrows are magnified by a factor 2.5.}
\label{fig:figure6}
\end{figure}

So far we have unveiled the main features of the crowd's mechanical response to a 
standardized perturbation, created by an intruder significantly larger than a pedestrian.
It is not  obvious  \emph{a priori} whether our findings apply to the case
where the intruder itself is a single pedestrian.
Therefore, we performed a second set
of experiments, in which single pedestrians crossed the static crowd, one by one.

To begin with, let us underscore the disparity between these new perturbations and those investigated above: Unlike the cylinder, crossing pedestrians are not of circular shape and have a much smaller cross section. In addition, while traversing they rotate their chest so as to `squeeze' through inter-pedestrian gaps in the crowd, which enhances the difference in shape and size. Finally, instead of walking straight, they followed winding trajectories, 
presumably favouring regions of lower density and larger spacings between pedestrians (some examples of trajectories are shown in Supplemental Fig.~S3).
This is consistent with the observation made in numerical simulations of a similar setting~\cite{metivet2018push} that making use of the empty spaces at the cost of a longer trajectory is favorable when the surrounding crowd is very dense.

These differences between the cylindrical and pedestrian intruders naturally affect the crowd's response. Because of their
smaller cross sections, the latter hardly alter the density of fairly
sparse crowds. For much
denser crowds, however, some pattern emerges in the perturbed density field,
shown in Fig.~\ref{fig:figure6}(a), with a clear dip in density in the wake of
the intruder. On the other hand, we find no evidence of the `wing'-like structure observed with the cylinder,
but considerably more data would be needed to make a definite statement in this respect.  Technically, it should nonetheless be mentioned that the winding
intruder's trajectory may add a `motion blur' along the $x$-coordinate to the density maps.

The mean velocity fields plotted in Fig.~\ref{fig:figure6}(c-d) give sharper insight into the induced perturbation. Remarkably, albeit noisier, the flow patterns bear vivid resemblance with those induced by the cylinder, even at fairly low density. Indeed, anticipated lateral moves are seen ahead of the intruder, while transverse inwards displacements fill the depleted region in its wake. Furthermore, the displacements are virtually all in the transverse direction. Recall that these are on no account trivial features, since they are not present in granular media.

The displacement amplitudes $\boldsymbol{\Delta u}$ [an example of which is
plotted in Fig.~\ref{fig:figure6}(b)] reflect both the similarity in the
crowd's response and the difference in the perturbation: For sparse crowds
(data not shown), the pedestrian-induced perturbation is too weak and no
clear profile emerges, but at higher density the profile of
$\boldsymbol{\Delta u}(x)$ resembles that observed with the cylindrical
intruder, with a fast
transverse decay. Two subtle differences
may nevertheless be pointed out. Firstly, the peak amplitude is only slightly
above 20 cm, consistently with the pedestrian's smaller `cross section'.
Secondly, the longitudinal displacement amplitude $\Delta u_y$ is not always
significantly smaller than $\Delta u_x$, possibly because of the winding
trajectory of the crossing pedestrian. 

All in all, the responses to the cylinder's and pedestrian's crossings share several common features. This qualitative robustness is encouraging for a prospective development of continuum crowd mechanics, based on empirical data rather than postulated rules.

\section*{Discussion}

\subsection*{Comparison with the response of granular media}

Let us now discuss whether our observations on pedestrian crowds crossed by a cylindrical intruder match previous findings for granular media. On the one hand, the gradually refilled depletion  zone in the intruder's wake mirrors the cavity formed behind the intruder in granular media. Besides,  we observed
perturbations that decay quite fast (in the transverse direction), with little
sensitivity to the crowd's density. Similarly, in granular media, the
perturbation decays exponentially in space\cite{seguin2013experimental}. Recall that this stands in stark contrast
with the very long range effects encountered in incompressible viscous fluids\cite{kaplun1957low} and also in the linear response of compressible elastic media to a point 
force\cite{karimi2015elasticity}. This similarity points to
the importance of the granularity of the crowd\footnote{Note, however, that more generic explanations could also be put forward. For instance, in fluids, shear-thinning generically promotes localisation near the intruder \cite{tanner1993stokes}.}.

On the other hand,  some striking differences were also found between the crowd's response and that of granular media. Ahead of the intruder, the pedestrians' avoidance strategies lead to anticipated lateral moves prior to contact. As a result, 
the compacted region does not take the shape of a dome as in granular matter below jamming, but rather forms a wing-like structure on both sides of the intruder.
Still, the most striking difference between pedestrians and grains
concerns the displacement field. In the case of pedestrians, the displacements are almost exclusively directed laterally (outwards or inwards), at odds with the loop-like (`dipolar') pattern with recirculation eddies seen in grains. It is remarkable that this feature survives when the intruder
is a single moving pedestrian.

Two specificities of pedestrians contribute to the realization
of such lateral displacements. The first one is the ability to \emph{anticipate}, which is reflected by the fact
that pedestrians start to move much before contact. This ability also
allows static pedestrians to forecast that they only need to move
laterally and let the intruder pass.
Incidentally, anticipation is so anchored in the pedestrians' behaviour that,  even when they are told not to anticipate, they still 
have some tendency to do so (provided they see the intruder coming).
The second specificity of pedestrians is \emph{self propulsion};
 they can  move in a direction which is not aligned
with the force applied on them, unlike grains.

Only when anticipation was precluded by the given instructions and the crowd's orientation
(with an intruder coming from behind) did the pedestrian displacement field turn more similar to that of grains,
whose motion results from contact forces. A sensible hypothesis is that, in the absence of anticipation,
the pedestrians are simply shoved by contact forces which they cannot withstand, while that they lack information to determine what moves would be optimal.

\subsection*{Comments and perspectives}

The foregoing differences in the response of pedestrians as compared to grains (whose origin we ascribed to anticipation and self-propulsion)
cast doubt on the the incautious use
of force models, even at densities where physical
forces are commonly believed to dominate. In particular, the observed prominence
of transverse displacements cannot be reproduced by models
based on radial interpedestrian \emph{forces}. Our findings would rather lend credence to 
descriptions based on (contact avoidance) \emph{strategies}.

More generally, the collection of experimental results on high density
crowds should foster the development of data-driven modelling efforts.
These efforts are encouraged by our finding of robust characteristics in the crowd's response,
which emerge in spite of the randomness stemming from the pedestrians' free will and their
diverse body sizes, and resist variations in the crowd's density.
These characteristics did not require extensive averaging (less than ten realisations).
Nonetheless, it would be desirable to extend our work to larger crowds and
to vary the intruder's size. Besides, the response to the crossing
of a single pedestrian deserves further study. Indeed, in this case, unlike that of the cylinder, the static pedestrians may expect
the intruder to also take his share of the efforts associated with the crossing,
by adapting his trajectory. How this possibly impacts the crowd's response is still unclear.

\section*{Methods}

\subsection*{Experiments}
The experiments were approved by the local
ethics committees (C3E and Grupo de Higiene y Seguridad) prior to their realisation and
they were conducted and analysed in accordance with the relevant guidelines and regulations. Informed 
consent was obtained from all participants.

\subsection*{Voronoi-based local densities with reduced edge effects}

\begin{figure}[ht]
\centering
\includegraphics[width=0.4\linewidth]{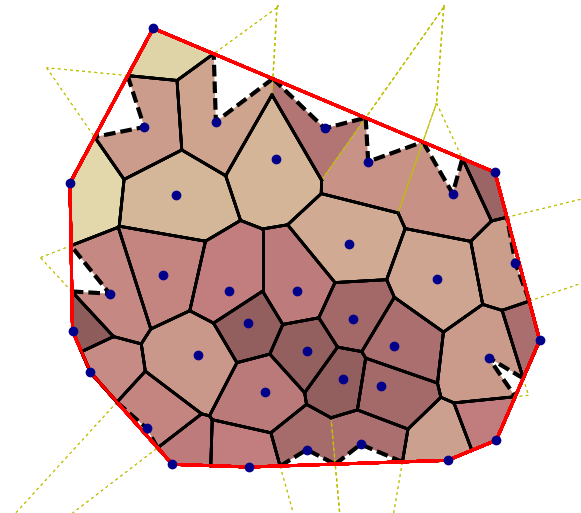}
\caption{Modified Voronoi tesselation used to compute local densities. The border $\mathcal{B}$ of the convex hull of the group
of pedestrians (blue dots) is represented as a red line and delimits the modified Voronoi cells of border points. Besides, the ridges of the initial Voronoi cells that extend
out of the convex hull (dotted yellow lines) are replaced by the dashed black lines. The solid and dashed segments in black (together with the convex hull) delimit the final Voronoi cells. }
\label{fig:figure7}
\end{figure}

The Voronoi tessellation is a convenient tool to delimit the area `belonging' to each pedestrian; local densities $\rho$ are then defined as the inverse area 
$\mathcal{A}^{-1}$ of the local Voronoi cell. But this definition fails for points (pedestrians) at the edge of the system, namely, (i) the points $P_B$ that form part of the border $\mathcal{B}$ of the convex hull of the system and (ii) the points $P_V$ whose Voronoi cells spread beyond the convex hull. These points are not surrounded in all directions and their personal space is thus bounded only in a portion of the spatial directions, which we call the populated sector. More precisely, the populated sector of a point $P_B$ is the angular sector limited by the two adjacent points in $\mathcal{B}$, while the populated sector of a point $P_V$ is delimited by the intersections between their Voronoi cell and $\mathcal{B}$, as shown in Fig.~\ref{fig:figure7}. For all edge points, the local density is the density seen in the populated sector (which forms an angle $\alpha$), i.e.,
 \begin{equation*}
\rho^{\prime}=\frac{\alpha}{2\pi} \frac{1}{\mathcal{A}^\prime},
 \end{equation*}
where $\mathcal{A}^\prime$ is the area of the Voronoi cell that lies in the populated sector.

Mean values for the local densities are obtained by averaging areas (i.e., $1/\rho^{\prime}$, and not $\rho^{\prime}$) over time and over realisations.

\subsection*{Smoothing of the displacement and velocity fields}

To get smoother (continuous) displacement fields out of our finite data set, we perform a convolution of the displacements $\boldsymbol{u}_{i}$ of all pedestrians $i$ as follows
\begin{equation}
\boldsymbol{u}(\boldsymbol{r},t)\equiv\frac{\sum_{i}\boldsymbol{u}_{i}(t)\,\phi(\left\Vert \boldsymbol{r}-\boldsymbol{r}_{i}(t)\right\Vert )}{\sum_{i}\phi(\left\Vert \boldsymbol{r}-\boldsymbol{r}_{i}(t)\right\Vert )},
\end{equation}
where $\phi$ is a smooth function (almost everywhere) with a compact support. Here, we have chosen ($\xi=0.25\,\mathrm{m}$), 
\begin{eqnarray}
\phi(r)=\begin{cases}
\exp\left(-\frac{r^{2}}{2\xi^{2}}\right) & \text{ if }r<2\xi\\
0 & \text{ otherwise}
\end{cases}
\label{eq:phi}
\end{eqnarray}

The continuous velocity field is obtained from the displacement field over
a time interval $\delta t$ as 
 $\boldsymbol{v}(\boldsymbol{r},t)=\frac{\boldsymbol{u}(\boldsymbol{r},t)}{\delta t}$
 
 \subsection*{Displacement amplitudes}
 
The displacement amplitude $\boldsymbol{\Delta u}^{(j)}$ of a pedestrian $j$ is defined by Eq.~\eqref{eq:deltau}. In order to get a smooth transverse profile
$\boldsymbol{\Delta u}(x)$, we compute the following convolution 
 with the coarse-graining function $\phi$
introduced in Eq.~\eqref{eq:phi}, viz.,
\begin{equation}
\boldsymbol{\Delta u}(x)= \frac{\sum_j \phi(x_j - x)\, \boldsymbol{\Delta u}^{(j)} }{ \sum_j \phi(x_j - x) },
\end{equation}
where $x_j$ denotes the transverse ($x$) distance between the initial pedestrian's position
and the intruder's trajectory, i.e., $x(t_i)-x_{cyl}(t_c)$ where $t_c$ is such that 
$y_{cyl}(t_c)=y(t_i)$.

\section*{Acknowledgements}
We thank all participants for taking part in the experiments.
A.N. and C.A.-R. acknowledge help from Tom Marzin, \'Etienne Pinsard, Aymeric Duigou-Majumdar, and Ioannis Touloupas for the organisation of the experiments
A.N. and C.A.-R. acknowledge funding from PALM  (ANR-10-LABX-0039-PALM) through the project PERCEFOULE.  M.K., S.I., and S.B. acknowledge help from A. Kolton and from the staff of Grupo de F\'isica Estad\'istica e interdisciplinaria (CAB)

\section*{Authors' contributions}

A.N. conceived the experiment.  A.N. and C.A.-R. conducted the French experiments, M.K., S.I., and S.B. conducted the Argentinian experiments. A.N. analysed the results of both experiments, in collaboration with C. A.-R..  A.N. and C.A.-R. wrote the manuscript. All authors reviewed the manuscript. 

\section*{Additional information}
The authors declare no competing interests.
The raw trajectories collected in relation with the experiments described here and the scripts used for the analysis are available upon request to the authors.

\end{document}